\documentclass[preprintnumbers,floatfix,12pt,superscriptaddress,nofootinbib]{revtex4}

\usepackage{amsfonts}
\usepackage{amssymb,epsfig}
\usepackage{latexsym}
\usepackage{amsmath}
\usepackage{graphicx}

\begin{document}

\title{ Cosmography of interacting generalized QCD ghost dark energy}
\author{Mohammad Malekjani \footnote{Email: \text{malekjani@basu.ac.ir}}}

 \address{Department of Physics, Faculty of Science, Bu-Ali Sina University, Hamedan 65178,
Iran}
\address{Research Institute for Astronomy and Astrophysics
of Maragha (RIAAM), Maragha,  P. O. Box:55134-441, Iran\\}

\begin{abstract}
Exploring the accelerated expansion of the universe, we investigate
the generalized ghost dark energy (GGDE) model from the statefinder
diagnosis analysis in a flat FRW universe. First we calculate the
cosmological evolution and statefinder trajectories for
non-interacting case and then extend this work by considering the
interaction between dark matter and dark energy components. We show
that in the non-interacting case the phantom line can not be crossed
and also he evolutionary trajectories of model in $s-r$ plane can
not be discriminated. It has been shown that the present location of
model in $s-r$ plane would be close to observational value for
negative values of model parameter. In the presence of interaction
between dark matter and dark energy, the phantom regime is achieved,
the accelerated phase of expansion occurs sooner compare with
non-interacting case. The GGDE model is also discussed from the
viewpoint of perturbation theory by calculating the adiabatic sound
speed of the model. Finally, unlike the non-interacting case, the
evolutionary trajectories in $s-r$ plane can be discriminated in the
interacting model. Like non-interacting model, in the interacting
case the present location of GGDE model is closer to observational
value for negative values of model parameter.
\end{abstract}
\maketitle

\newpage
\section{Introduction}
The recent astronomical data from SNe Ia \cite{c1}, WMAP \cite{c2},
SDSS \cite{c3} and X-ray \cite{c4} experiments show that our
universe experiences an accelerated expansion. The above
observational data strongly suggest that the universe is spatially
flat and dominated by an exotic component with negative pressure,
the so-called dark energy
\cite{winberg1,winberg2,winberg3,winberg4}. Dark energy scenario has
got a lot of attention in modern cosmology both from theoretical and
observational point of view. Observationally, The result of WMAP
experiment shows that dark energy occupies about $73\%$ of the
energy of our universe, dark matter about $23\%$ and usual baryons
occupies only about $4\%$ of the total energy of the universe {c2}.
Although the nature of dark energy is still un-known, but the
ultimate fate of the current universe is determined by this
mysterious component. Theoretically, the first and simplest model
for dark energy is Einstein's cosmological constant with constant
EoS parameter $w_{\Lambda}=-1$ . Cosmological constant faces with
the fine-tuning and cosmic coincidence problems
\cite{fin,winberg1,winberg2,winberg3,winberg4,reva,revb,revc,revd,reve,revf,revg}.
In recent years a plenty theoretical models have been proposed to
interpret the properties of dark energy \cite{reva,revb,revc,revd,reve,revf,revg,rev2,rev3}.\\
Almost all of theoretical dark energy models need to introduce new
degree(s) of freedom or modifying general relativity. However it
would be better to consider a model of dark energy without a need of
new degree(s) or new parameter in its theory. Recently the so-called
QCD ghost dark energy has been proposed to interpret the dark energy
without any new parameter or new degree of freedom
\cite{ghostq1,ghostq2,ghostq3,ghostq4,ghostq5,ghostq6,ghostq7}. The
Veneziano ghost has been suggested to solve the U(1) problem in low
energy effective theory of QCD
\cite{witq1,witq2,witq3,witq4,witq5,witq6}. The ghost field has no
contribution to the vacuum energy density in the flat Minkowsi
spacetime. However, in the case of curved spacetime it has a small
energy density proportional to $\Lambda_{QCD}^3H$, where
$\Lambda_{QCD}$ is QCD mass scale and $H$ is the Hubble parameter
\cite{zhiten1,zhinten2,zhinten3}. This model does not encounter with
some unwanted problems such as the violation of gauge invariance,
unitarity and causality \cite{zhiten1}. Since the Veneziano ghost
field is totally embedded in standard model and general relativity,
one needs not to introduce any new degree(s) of freedom or to modify
the Einstein's general relativity. The present value of energy
density of dark energy in this model is roughly of order
$\Lambda_{QCD}^3H_0$, with $\Lambda_{QCD}\sim100 MeV$ and
$H_0\sim10^{-33}eV$ which is in agreement with observed value
$(3\times10^{-3}eV)^4$ for energy density of dark energy
\cite{caiq1}. This numerical coincidence is adsorbent and gets rid
the model from fine tuning problem
\cite{ghostq1,ghostq2,ghostq3,ghostq4,ghostq5,ghostq6,ghostq7}.
Observationally, the ghost dark energy model has been fitted by
astronomical data including SnIa, BAO, CMB, BBN and Hubble parameter
data \cite{ghostq22}. The cosmological evolution of dark energy in
the QCD ghost model has been calculated in \cite{ghostq1,ghostq2}
and has been resulted that the universe begins to accelerate at
redshift around $z\sim 0.6$. Also the squared sound speed of the
dark energy for this model is negative, indicating an instability of
the model against perturbational theory
\cite{ghostq1,ghostq2,ghostq3,ghostq4}. In \cite{sheykhi1}, the
ghost dark energy (GDE,hereafter) has been extended in the presence
of interaction between dark matter and dark energy in non-flat
universe. The reconstructed potential and the dynamics of scalar
fields according the exultation of GDE have been investigated in
\cite{sheykhi2a,sheykhi2b,sheykhi2c,sheykhi2d,sheykhi2e}. The
reconstructed modified gravity for GDE which describes the late time
accelerated expansion has been studied in \cite{khod1}. The
statefinder diagnostic of GDE has been presented in \cite{malek1}.
Form the statefinder viewpoint, the evolution of GDE model is
similar to holographic dark energy model and present value of
statefinder parameters in this model is in good agreement
with observation \cite{malek1}.\\
In all above studies, the energy density of GDE is considered
proportional to Hubble parameter as $\rho_d=\alpha H$. However, the
energy density of Veneziano ghost field in QCD is generally in the
form of $H+O(H^2)$ \cite{zhit}. In this case the $U(1)_A$ problem in
QCD can be solved. Although, up to now, only the leading term $H$
has been assumed for energy density of GDE, but the sub-leading term
$H^2$ can also be important in the early evolution of the universe
\cite{h2}. Including the second term in the energy density of GDE
results better agreement with observation in comparison with usual
GDE model \cite{caighost}. Like \cite{sheykhi3}, we call this model
as generalized ghost dark energy (GGDE). The energy density of GGDE
model is written as $\rho_d=\alpha H+\beta H^2$, where $\alpha$ and
$\beta$ are the constants of the model. It has been shown that the
GGDE model can result a de-Sitter phase of expansion and also in the
presence of interaction between dark matter and dark energy this
model results the phantom regime of expansion ($w_d<-1$)
\cite{sheykhi3}. The other features of GGDE model have been
presented in
\cite{sheykhi4, sheykhib}.\\
It is well known that in addition of dark energy component which
describe the accelerated expansion of the universe, there exist
another mysterious component in the universe so-called dark matter.
The dark matter component can interpret the flat rotation curve of
spiral galaxies and also the scenario of structure formation of
universe \cite{DM, DM2, DM3, DM4, DM5}. Since the nature of these
component are un known and they they have different gravitational
treatment, therefore their evolution usually considered independent
of each other. However recent observation from galaxy cluster Abell
A596 indicates the interaction between these components
\cite{interact}. Also the observational data from SNIa and CMB
experiments is compatible with interacting forms of dark energy
models \cite{interact2}. However the strength of this interaction is
not clearly identified \cite{feng}. From theoretical viewpoint it is
also acceptable to consider the interaction between dark matter and
dark energy. In the unified models of field theory dark matter and
dark energy can be interpreted by a single scalar field in a
minimally interaction. Also considering
interaction between dark matter and dark energy can solve the coincidence problem \cite{new1,new2,new3,new4,new5,new6,new7,new8}  \\

In this work our main task is to investigate the interacting GGDE
model in statefinder diagnostic analysis. Different dynamical dark
energy models obtain accelerated expansion at the present time
($q<0$), where $q$ is deceleration parameter. Hence we need a
diagnostic tool for discriminating these dark energy models. For
this aim, Sahni et al. \cite{sah03} and Alam et al. \cite{alamb}, by
using the third time derivative of scale factor, introduced the
statefinder pair $\{s,r\}$. These parameters in flat universe are
given by

\begin{equation}\label{state1}
r=\frac{\dddot{a}}{aH^3},s=\frac{r-1}{3(q-1/2)}
\end{equation}

The parameters $s$ and $r$ are geometrical, because they only depend
on the scale factor. In statefinder analysis we plot the
evolutionary trajectories of dark energy model is $s-r$ plane. In
recent years the various dark energy models such as quintessence,
holographic,new holographic, phantom, tachyon, chaplygin gas,
agegraphic, new agegraphic, , polytropic gas and ghost dark energy
models have been studied in the statefinder analysis
\cite{stateref1,stateref2, stateref3, stateref4,
stateref5,stateref6,stateref7,stateref8,stateref9,stateref10,stateref11,stateref12,stateref13,malekghost,sah03,alamb}.
These models have different evolutionary trajectories in \{s, r\}
plane, therefore the statefinder tool can discriminate these models.
The standard $\Lambda$CDM has no evolution in this plane and
corresponds to the fixed point \{s=0,r=1\} \cite{sah03}. The present
observational value for statefinder parameters are $\{s_0=-0.006
,r=1.02\}$ \cite{kapa}. The distance of the current value of
statefinder pair $\{s_0, r_0\}$ of a given dark energy model from
the observational value $\{s_0=-0.006 ,r=1.02\}$ is a valuable
criterion to examine of model. Here we see that the location of
standard $\Lambda$CDM model in $s-r$ plane is near to observational
value. In \cite{malekghost}, the evolution of original GDE has been
calculated by statefinder diagnostic in $s-r$ plane and shown that
the GDE model mimics the $\Lambda$CDM at the late time. Also the
behavior of GDE is similar
to holographic dark energy in this plane \cite{malekghost}.\\
In this work we first calculate the cosmological evolution of GGDE
model and then investigate this model from statefinder diagnostic
analysis. The paper is organized as follows: In sect.II, The GGDE
model is presented in non-interacting universe. The interacting case
of GGDE model is given in sect.III. In sect. IV we obtain the
adiabatic sound speed for GGDE model. We calculate the numerical
results in sect.V and conclude in sect.VI.

\section{Non-Interacting GGDE model}
A flat Friedmann-Robertson-Walker (FRW) universe dominated by dark
matter and dark energy is given by
\begin{equation}\label{fridt}
H^{2}=\frac{1}{3m_{p}^{2}}(\rho _{m}+\rho _d)
\end{equation}%
where $\rho_m$ and $\rho_d$ are, respectively, the energy density of
pressureless dark matter and dark energy and $m_p$ is the reduced
planck mass. The energy density of GGDE is given by \cite{sheykhi3}
\begin{equation}\label{statet}
\rho_{\Lambda}=\alpha H+ \beta H^2
\end{equation}
where $\alpha$ and $\beta$ are constants of model. The Friedmann
equation (\ref{fridt}) in terms of dimensionless parameters is
written as
\begin{equation}
\Omega _{m}+\Omega _{\Lambda}=1.  \label{Freqt}
\end{equation}%
where
\begin{equation}\label{denergyt}
\Omega_{m}=\frac{\rho_m}{\rho_c}=\frac{\rho_m}{3M_p^2H^2}, ~~~\\
\Omega_d=\frac{\rho_d}{\rho_c}=\frac{\rho_d}{3M_p^2H^2}~~\\
\end{equation}
The conservation equations for pressureless dark matter and dark
energy without interaction read the following equations
\begin{eqnarray}
\dot{\rho _{m}}+3H\rho _{m}=0, \label{contmt}\\
\dot{\rho _d}+3H(1+w_d)\rho_d=0. \label{contdt}
\end{eqnarray}%
Taking the time derivative of Friedmann equation (\ref{fridt}) and
using (\ref{Freqt}, \ref{contmt}, \ref{contdt}) obtains
\begin{equation}\label{hdott}
\frac{\dot{H}}{H^2}=-\frac{3}{2}[1+w_{\Lambda}\Omega_d]
\end{equation}
Differentiating Eq.(\ref{statet}) with respect to time yields
\begin{equation}\label{dotdens}
\dot{\rho}=\dot{H}(\alpha+2\beta H)
\end{equation}

Inserting (\ref{dotdens}) and (\ref{statet}) in conservation
equation for dark energy (\ref{contdt}) and using (\ref{hdott}), the
EoS parameter of GGDE model can be obtained as
\begin{equation}\label{eos1}
w_d=\frac{\xi-\Omega_d}{\Omega_d(2-\Omega_d-\xi)}
\end{equation}
where $\xi=8\pi G\beta/3$. In the limiting case $\xi=0$, this
relation reduces to its original form in \cite{sheykhi1}. The
deceleration parameter $q$ by using (\ref{hdott}) and (\ref{eos1}),
in GGDE universe is obtained as
 \begin{equation}\label{q}
 q=\frac{1}{2}-\frac{3}{2}\frac{\xi-\Omega_d}{(\xi+\Omega_d-2)}
 \end{equation}
The decelerated phase of expansion at the early time is indicated by
$q<0$ and accelerated phase is related to $q>0$. Taking the time
derivative of dimensionless dark energy density in (\ref{denergyt})
and using (\ref{statet}), (\ref{dotdens}), we obtain the equation of
motion for the evolution of energy density of GGDE model as

\begin{equation}\label{omegp}
\Omega_d^{\prime}=-3\frac{(1-\Omega_d)(\xi-\Omega_d)}{(2-\Omega_d-\xi)}
\end{equation}
where prime is derivative with respect to $\ln{a}$. Taking a
derivative of (\ref{eos1}) with respect to $\ln{a}$, the equation of
motion for EoS parameter can be calculated as
\begin{equation}\label{wp}
w_d^{\prime}=\frac{3(1-\Omega_d)(\xi-\Omega_d)}{\Omega_d(2-\Omega_d-\xi)^2}\Big[1+\frac{(\xi-\Omega_d)(2-2\Omega_d-\xi)}{\Omega_d(2-\Omega_d-\xi)}\Big]
\end{equation}
Using the above relation, in this stage, we calculate the
statefinder parameters $s$ and $r$ for GGDE model in non-interacting
universe. In general form, relation (\ref{state1}) for a given dark
energy model in flat universe can be written as
\begin{equation}\label{r1}
r=1+\frac{9}{2}w_d\Omega_d(1+w_d\Omega_d)-\frac{3}{2}(w_d^{\prime}\Omega_d+w_d\Omega_d^{\prime})
\end{equation}
and
\begin{equation}\label{s1}
s=1+w_d\Omega_d-\frac{1}{3}(\frac{w_d^{\prime}}{w_d}+\frac{\Omega_d^{\prime}}{\Omega_d})
\end{equation}
Inserting relations (\ref{omegp}) and (\ref{wp}) in equations
(\ref{r1}) and (\ref{s1}), we obtain the statefinder parameters for
GGDE model in spatially flat universe

\begin{equation}\label{r2}
r=1+9\frac{(\xi-\Omega_d)(1-\Omega_d)^2}{(2-\Omega_d-\xi)^3}
\end{equation}

\begin{equation}\label{s2}
s=\frac{2(1-\Omega_d)^2}{(2-\Omega_d-\xi)^2}
\end{equation}
In the limiting case of dark energy dominated universe
($\Omega_d\rightarrow 0$) the parameters \{s,r\} tends to $\{0,1\}$,
respectively. Hence the GGDE model mimics the $\Lambda$CDM model at
the late time when $\Omega_d \rightarrow 0$. In sect.V, we calculate
numerically the evolution of GGDE model in non-interacting universe
from the statefinder viewpoint.

\section{Interacting GGDE model}
In this section we consider the interaction between dark matter and
dark energy components. In this case the conservation equations for
these components are:
\begin{eqnarray}
\dot{\rho _{m}}+3H\rho _{m}=Q, \label{icontmt}\\
\dot{\rho _d}+3H(1+w_d)\rho_d=-Q. \label{icontdt}
\end{eqnarray}%
where $Q$ in right hand side indicate the interaction term. The
positive value of $Q$ means the transition of energy from dark
energy to dark matter component. It should be noted that the left
side of (\ref{contmt}) and (\ref{contdt}) are inversely proportional
to time. Therefore the parameter $Q$ can be considered as a function
of Hubble parameter $H$ such as following forms:\\
 (i) $Q\propto H\rho_{d}$\\ (ii) $Q\propto H\rho_{m}$ \\ (iii) $Q\propto
H(\rho_{m}+\rho_{d})$.\\ One can assume the above three forms as
$Q=\Gamma \rho_{d}$, where for case (i) $\Gamma=3b^2H$, for case
(ii) $\Gamma=3b^2H \frac{\Omega_m}{\Omega_{d}}$ and for case (iii)
$\Gamma=3b^2H \frac{1}{\Omega_{d}}$. The parameter $b$ is a coupling
constant indicating the strength of interaction
between dark matter and dark energy \cite{c21,c21b,c21c}. In this work we assume the third form of interaction (i.e., $Q=3Hb^2\frac{\rho_d}{\Omega_d}$).\\
Substituting $Q$ in (\ref{icontdt}) and using (\ref{statet}),
(\ref{hdott}) and (\ref{dotdens}), the EoS parameter of interacting
GGDE model is obtained as
\begin{equation}\label{eos2}
w_d=\frac{\xi-\Omega_d-2b^2}{\Omega_d(2-\Omega_d-\xi)}
\end{equation}
Inserting $b=0$ recovers the EoS parameter of non-interacting case
in previous section. Substituting (\ref{eos2}) in (\ref{hdott})
results the deceleration parameter $q$ for interacting case as
follows
 \begin{equation}\label{iq}
 q=\frac{1}{2}-\frac{3}{2}\frac{(\xi-\Omega_d-2b^2)}{(\xi+\Omega_d-2)}
 \end{equation}
It has been shown that for selected parameters ($\xi=0.03, b=0.15,
\Omega_{d0}=0.72$) the deceleration parameter at the present time is
$q_0=-0.38$ which is consistent with observation \cite{sheykhi3}.\\
Taking the time derivative of dimensionless dark energy density in
(\ref{denergyt}) and using (\ref{statet}), (\ref{dotdens}) and
(\ref{eos2}) , the equation of motion for the evolution of energy
density of interacting GGDE model can be obtained as

\begin{equation}\label{iomegp}
\Omega_d^{\prime}=-3\Big[\frac{(1-\Omega_d)(\xi-\Omega_d-2b^2)}{(2-\Omega_d-\xi)}+b^2\Big]
\end{equation}
In the limiting non-interacting case ($b=0$) the respective relation
in previous section is retained. Derivative of (\ref{eos2}) with
respect to $\ln{a}$ results
\begin{equation}\label{iwp}
w_d^{\prime}=\frac{3(1-\Omega_d)(\xi-\Omega_d-2b^2)+b^2(2-\Omega_d-\xi)}{\Omega_d(2-\Omega_d-\xi)^2}\Big[1+\frac{(\xi-\Omega_d-2b^2)(2-2\Omega_d-\xi)}{\Omega_d(2-\Omega_d-\xi)}\Big]
\end{equation}
Once again, the respective relation in previous section can be
obtained by setting $b=0$. Finally by substituting relations
(\ref{iomegp}) and (\ref{iwp}) in general relations (\ref{r1}) and
(\ref{s1}), we obtain the statefinder parameters for interacting
GGDE model as follows
\begin{equation}\label{r3}
r=1+9\frac{(\xi-\Omega_d-2b^2)(1-\Omega_d-b^2)}{(2-\Omega_d-\xi)^2}-9\frac{(1-\Omega_d)(\xi-\Omega_d-2b^2)+b^2(2-\Omega_d-\xi)}{(2-\Omega_d-\xi)^3}(1-\xi+b^2)
\end{equation}

\begin{equation}\label{s3}
s=2\Big[\frac{(1-\Omega_d-b^2)}{(2-\Omega_d-\xi)}-\frac{(1-\Omega_d)+b^2\frac{2-\Omega_d-\xi}{\xi-\Omega_d-2b^2}}{(2-\Omega_d-\xi)^2}(1-\xi+b^2)\Big]
\end{equation}
Setting $b=0$ retains the relations for $s$ and $r$ in previous
section. In section V we investigate the evolution of interacting
GGDE model in $s-r$ plane can calculate the effect of interaction
parameter $b^2$ on the evolution of the model.

\section{Adiabatic sound speed}
In linear perturbation theory, squared sound speed, $c^2$ is a
crucial quantity. Stability or instability of a given perturbed mode
can be calculated by determining the sign of $c^2$. The positive
sign (real value of sound speed) represents the periodic propagating
mode for a density perturbation and in this case we have the
stability. The negative sign (imaginary value of sound speed)
indicates an exponentially growing mode for a density perturbation,
meaning the instability \cite{sound1,sound2}. Here we obtain the
squared sound speed for GGDE model both in non-interacting and
interacting cases. The squared sound speed $c_s^2$ is introduced as
\begin{equation}\label{sou1}
c_s^2=\frac{dp}{d\rho_d}=\frac{\dot{p}}{\dot{\rho_d}}
\end{equation}
We now differentiate the equation of state, $p_d=w_d\rho_d$ with
respect to time and find
\begin{equation}\label{sou2}
\dot{p_d}=\dot{w_d}\rho_d+w_d\dot{\rho_d}
\end{equation}
Inserting (\ref{sou2}) in (\ref{sou1}) and using Eq.(\ref{contdt}),
we obtain $c_s^2$ for non-interacting GGDE model as follows
\begin{equation}\label{sou3}
c_s^2=w_d-\frac{w_d^{\prime}}{3(1+w_d)}
\end{equation}
where prime is the derivative with respect to $\ln{a}$ and
$w_d^{\prime}=\dot{w_d}/H$. In the case of interacting GGDE model by
using Eq.(\ref{icontdt}), the parameter $c_s^2$ can be obtained as
\begin{equation}\label{sou4}
c_s^2=w_d-\frac{w_d^{\prime}}{3(1+w_d)+\frac{3b^2}{\Omega_d}}
\end{equation}
Here same as previous section we used third form of interaction
parameter $Q=3Hb^2\rho_d/\Omega_d$. In next section, we obtain the
evolution of $c_s^2$ as a function of cosmic redshift and discuss
the stability or instability of GGDE model for both non-interacting
and interacting universe.

\section{Numerical results}
Here we present numerical description for cosmological evolution and
statefinder analysis of GGDE model in the flat FRW cosmology. In
numerical procedure we fix the cosmological parameters at the
present time as $\Omega_m^0=0.3$ and $\Omega_d^0=0.7$. We first
consider non-interacting case and then interacting case of GGDE
model.
\subsection{non-interacting case}

The EoS parameter of non-interacting GGDE model as a function of
density parameter $\Omega_d$ is given by (\ref{eos1}). By solving
coupled equations (\ref{omegp}) and (\ref{eos1}), the evolution of
EoS parameter in terms of cosmic redshift $z=1/a-1$ and for
different illustrative values of $\xi$ is shown in Fig.(1). The
cosmic redshift $z=0$ represents the present time, $z>0$ indicates
the past times and $z<0$ expresses the future. We see that for any
value of $\xi$ the non-interacting GGDE model can not enter the
phantom regime ($w_d<-1$) at all. We also see that the EoS of GGDE
model with $\xi>0$ is larger than EoS of standard GDE model
($\xi=0.0$). At the future epoch, the EoS tends to $-1$ which
implies that the GGDE model mimics the cosmological constant at that
time.\\

The deceleration parameter $q$ which indicates the decelerated or
accelerated phase of expansion for non-interacting case is given by
(\ref{q}). Solving coupled equations (\ref{q}) and (\ref{omegp}),
the evolution of parameter $q$ as a function of redshift parameter
$z$ has been shown in Fig.(2) for different values of model
parameter $\xi$. The standard GDE model is indicated by solid line.
one can see that for $\xi<0$ the GGDE model enters the accelerated
phase sooner and for $\xi>0$ later compare with standard GDE
model.\\

The statefinder pair \{s,r\} for non-interacting GGDE model is given
by relations(\ref{r2}) and (\ref{s2}). Solving these coupled
equations together with (\ref{omegp}), we obtain the evolution of
parameters $s$ and $r$ in terms of redshift $z$. In Fig.(3), we plot
the evolutionary trajectories of non-interacting GGDE model for
different values of model parameter $\xi$ in $s-r$ plane. We see
that the parameter $r$ first decreases and then increases and also
the parameter $s$ decreases during the history of the universe from
past to future. The important note is that in non-interacting case
the GGDE model has been shown by single evolutionary trajectory  for
any value of $\xi$. Hence the evolutionary trajectories can not
discriminated by model parameter $\xi$. The present value of
statefinder pair $\{s_0,r_0\}$ is indicated by colored circle on the
figure. Also the location of $\Lambda$CDM model in $s-r$ plane
,i.e., $(s=0,r=1)$, has been shown by star symbol. The other feature
is that the present value  $\{s_0,r_0\}$ is discriminated by model
parameter $\xi$. In the case of $\xi<0$, the distance of
$\{s_0,r_0\}$ from observational point $\{s_0=-0.006 ,r=1.02\}$ (red
star point on the figure) is shorter compare with standard GDE model
(i.e., $\xi=0.0$). While for $\xi>0$ the distance is larger than GDE
model. Finally we discuss numerically the stability or instability
of non-interacting GGDE model form the viewpoint of pertrurbation
theory based on Eq.(\ref{sou3}). In Fig.(4), the evolution of
adiabatic sound speed $c_s^2$ is plotted as a function of redshift
$z$ for different values of GGDE model $\xi$. In the case of
$\xi\geq 0$, one can see $c_s^2<0$ which indicates the instability
of GGDE model against perturbation. For $\xi<0$, we obtain $c_s^2>0$
which represents the stability of model against perturbation.

\subsection{interacting case}

Here we calculate the numerical description of cosmological
evolution and statefinder diagnosis for interacting GGDE model in
spatially flat universe. First the evolution of EoS parameter in
terms of cosmic redshift is plotted in Fig.(4). For this aim we
solved the coupled equations (\ref{eos2}) and (\ref{iomegp}). In
left panel, by fixing interaction parameter as $b=0.2$, the EoS
parameter $w_d$ is plotted for different illustrative values of
model parameter $\xi$ as described in legend. The solid line
represents the original GHDE model. For all cases of $\xi$, the
interacting case of GGDE model can cross the phantom line ($w_d=-1$)
form upper limit($w_d<-1$) to lower limit ($w_d<-1$). This behavior
of interacting GGDE model in which the phantom line is crossed from
up to below is in agrement with recent observations \cite{obsegg}.
In right panel the model parameter is fixed as $\xi=0.1$ and the
interaction parameter $b$ is varied. We see that the non-interacting
GGDE model ($\b=0.0$) cannot enter the phantom regime (solid line),
while in other cases ($\xi\neq 0$) the phantom regime has been
achieved.\\

In Fig.(5), by numerical solving of relations (\ref{iomegp}) and
(\ref{iq}), the evolution of deceleration parameter in terms of
redshift has been shown in the context of interacting GGDE model. In
left panel, by fixing $b=0.2$, the parameter $\xi$ is varied as
indicated in legend. Same as non-interacting case, the parameter $q$
starts from positive value at the earlier ( representing the
decelerated phase at the past time) and ends to negative value later
(indicating the accelerated phase at the present time). Like
non-interacting case, transition from decelerated phase to
accelerated phase for $\xi<0$ takes place earlier and for $\xi>0$
later, compare with original GDE model (see left panel of Fig.(5)).
In right panel, for an illustrative value $\xi=0.1$, the evolution
of $q$ has been shown for different values of interaction parameter
$b$. The interaction parameter $b$ can influence on the transition
epoch from $q>0$ to $q<0$. We see that in the framework of
interacting GGDE model the accelerated phase of expansion ($q<0$)
can be achieved sooner for larger values of $b$. For all cases,
$q\rightarrow -1$ at the late time indicating that the GGDE model
mimics the standard $\Lambda$ CDM model at that time.\\

Now, by solving relations (\ref{s3}) and \ref{r3}), the evolution of
interacting GGDE model in $s-r$ plane is plotted in Fig.(6). In left
panel, by fixing $\xi=0.1$, the interaction parameter $b$ is varied
as indicated in legend. The evolutionary trajectories starts from
right to left and ended at the $\Lambda$CDM fixed point. The
distance of present value $\{s_0,r_0\}$ form the observational point
$\{s_0=-0.006 ,r=1.02\}$ (red star point on the figure) becomes
larger by increasing $b$. In right panel, by fixing $b=0.2$, the
trajectories have been plotted for different illustrative values of
$\xi$. The important note is that, contrary with non-interacting
case, in the presence of interaction between dark matter and dark
energy ($b\neq0$) the evolutionary trajectories in $s-r$ plane are
discriminated by parameter $\xi$. Also the distance of $\{s_0,r_0\}$
from observational point $\{s_0=-0.006 ,r=1.02\}$ is shorter for
$\xi<0$ and larger for $\xi>0$ compare with original GDE model
(i.e., $\xi=0.0$). Finally, same as non-interacting case, we
investigate the interacting GGDE model from the viewpoint of
perturbation theory. The adiabatic sound speed $c_s^2$ for
interacting case is given by Eq.(\ref{sou4}). In Fig.(8), we
calculate the evolution of $c_s^2$ as a function of cosmic redshift
for different values of model parameter $\xi$ as well as interaction
parameter $b$. In left panel, by fixing model parameter $\xi=0.5$,
we obtain the evolution of $c_s^2$ for different values of
interaction parameter $b$. Here one can interpret that same as
non-interacting case, the interacting GGDE model for $\xi>0$ is
instable ($c_s^2<0$) against perturbation. In right panel, by fixing
interaction parameter $b=0.1$, the evolution of $c_s^2$ has been
shown for different values of model parameter $\xi$. Like
non-interacting case, we conclude that the interacting GGDE model is
stable for $\xi<0$ and instable for $\xi\geq 0$.

\section{conclusion}
In summary, we considered the generalized version of QCD ghost dark
energy (GGDE) model in both non-interacting and interacting
universe. The cosmological evolution and also the statefinder
diagnosis of the model have been calculated. We showed that:\\
(i). In the non-interacting GGDE model the phantom regime can not be
achieved and the EoS parameter reaches to asymptotic value $w_d=-1$
at the late time. We also showed that for negative values of model
parameter $\xi$ the transition from decelerated to accelerated phase
takes place sooner compare with original ghost dark energy (GDE)
model. The statefider analysis was also performed for
non-interacting GGDE model. In this case, in the absence of
interaction between dark matter and dark energy, the evolutionary
trajectories of model in $s-r$ plane can not discriminated. However,
the present value of statefinder parameter $\{s_0,r_0\}$ of the
model is diagnosed by parameter $\xi$. We concluded that
$\{s_0,r_0\}$ is closer to observational value $\{s_0=-0.006
,r=1.02\}$ for negative values of $\xi$ (see Fig.(3)). We also
obtained the stability or instability of the model against
perturbation by calculating the adiabatic sound speed and showed
that the non-interacting case of GGDE model is instable for $\xi\geq 0$ and stable for $\xi<0$\\
(ii). In the presence of interaction between dark matter and dark
energy, the GGDE model can cross the phantom line from up to down in
agreements with observations \cite{obsegg1,obsegg2,obsegg3}. Also,
in the context of interacting GGDE model the entrance to accelerated
phase of expansion occurs earlier compare with non-interacting case.
The statefinder diagnosis analysis was also performed for
interacting case and we showed that in the presence of interaction
the evolutionary trajectories of GGDE model $s-r$ plane are
diagnosed. Same as non-interacting model, the present value
$\{s_0,r_0\}$ is closer to observational point for negative values
of $\xi$ (see Fig.(6)).
We also showed that in the presence of interaction, the GGDE model has stability for $\xi<0$ and instability for $\xi\geq 0$.\\

\noindent{{ Acknowledgements}}\\
This work has been supported financially by Research Institute for
Astronomy $\&$ Astrophysics of Maragha (RIAAM) under research
project 1/2782-51.

\newpage

\newpage

\begin{figure}[!htb]
\includegraphics[width=8cm]{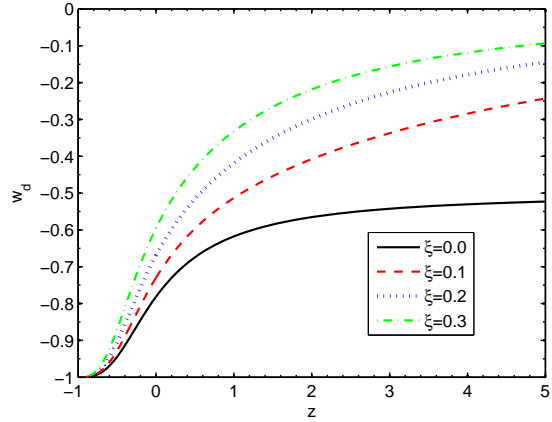}
\caption{The evolution of EoS parameter of non-interacting GGDE
model in terms of redshift parameter $z$ for different illustrative
values of model parameters $\xi$. Here we take $\Omega_d^0=0.70$ and
$\Omega_m^0=0.30$. }
\end{figure}

\begin{figure}[!htb]
\includegraphics[width=8cm]{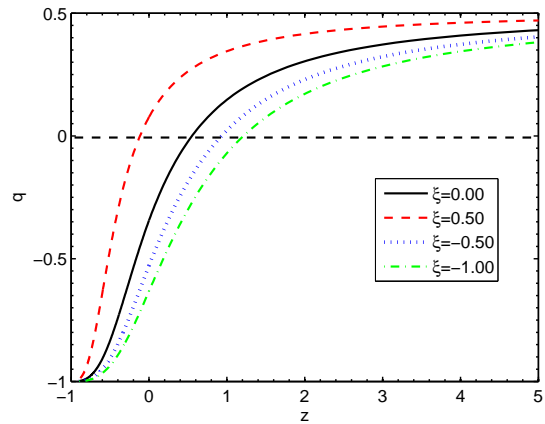}
\caption{The evolution of deceleration parameter $q$ in the context
of non-interacting GGDE model as a function of redshift $z$ for
different illustrative values of model parameters $\xi$. We toke
$\Omega_d^0=0.70$ and $\Omega_m^0=0.30$. }
\end{figure}

\begin{figure}[!htb]
\includegraphics[width=8cm]{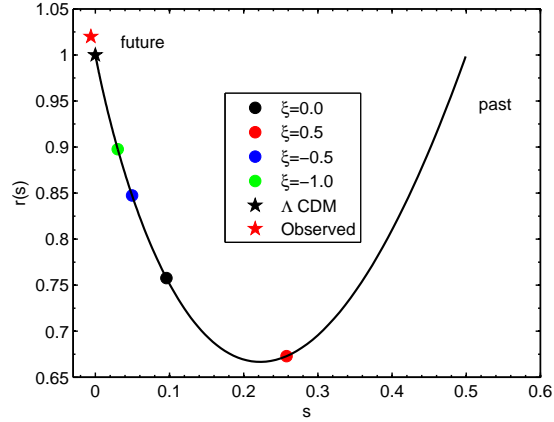}
\caption{The statefinder plot of non-interacting GGDE model in flat
FRW universe with $\Omega_d^0=0.70$ and $\Omega_m^0=0.30$. The
present value $\{s_0,r_0\}$ is indicated by colored circle on the
curves. The location of $\Lambda$CDM model and observational point
are indicated by black and red stars, respectively. }
\end{figure}

\begin{figure}[!htb]
\includegraphics[width=8cm]{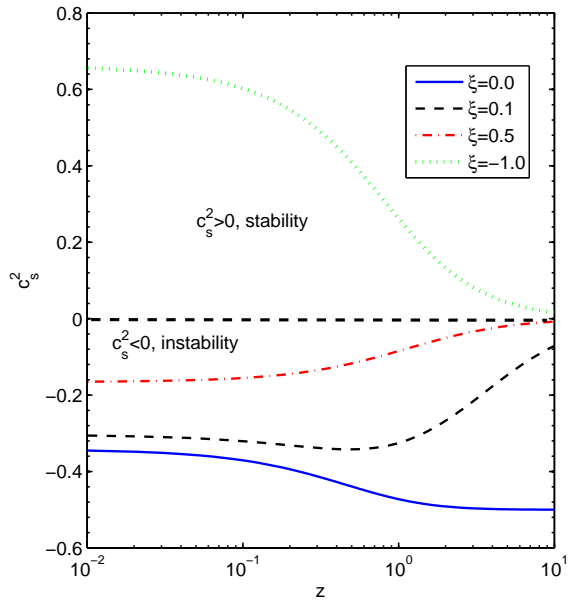}
\caption{The adiabatic sound speed $c_s^2$ as a function of cosmic
redshift $z$ for different model parameter $\xi$ as described in
legend. The horizontal dashed line separates the stability and
instability regions.}
\end{figure}

\begin{figure}[!htb]
\includegraphics[width=8cm]{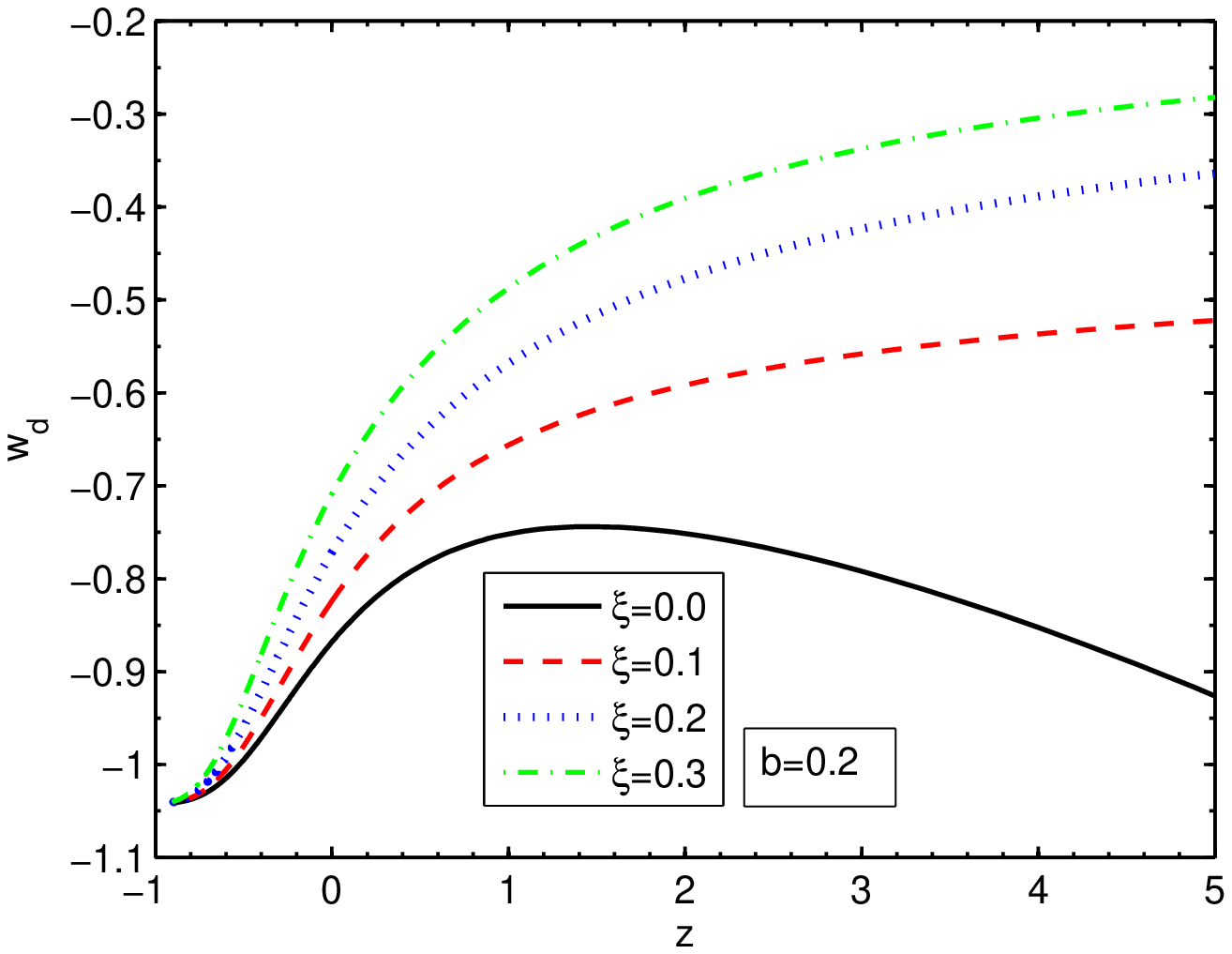}\includegraphics[width=8cm]{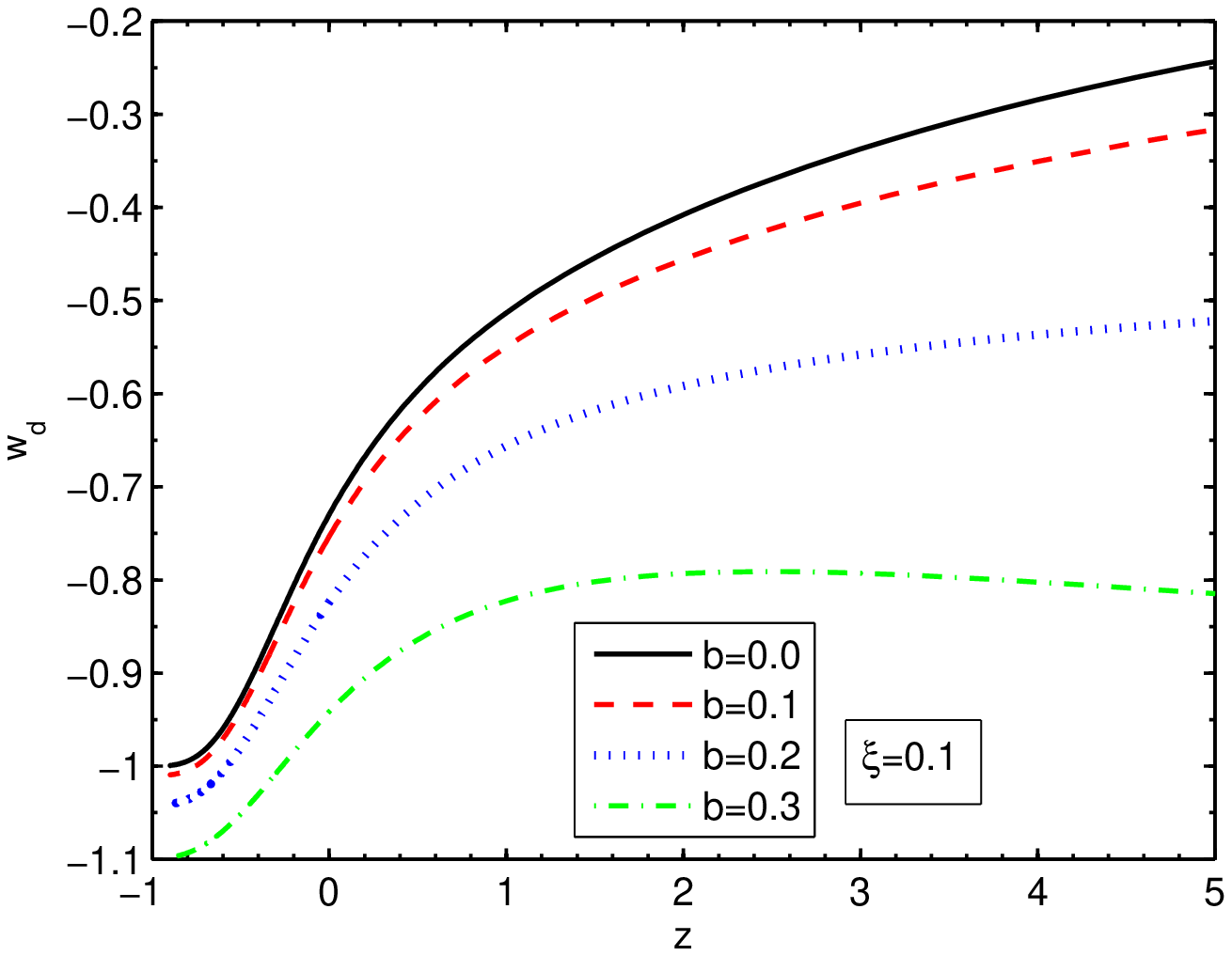}
\caption{The evolution of EoS parameter of interacting GGDE model
versus redshift parameter $z$ as described in legend. In left panel,
the interaction parameter $b$ is fixed and in the right panel the
parameter $\xi$. }
\end{figure}

\begin{figure}[!htb]
\includegraphics[width=8cm]{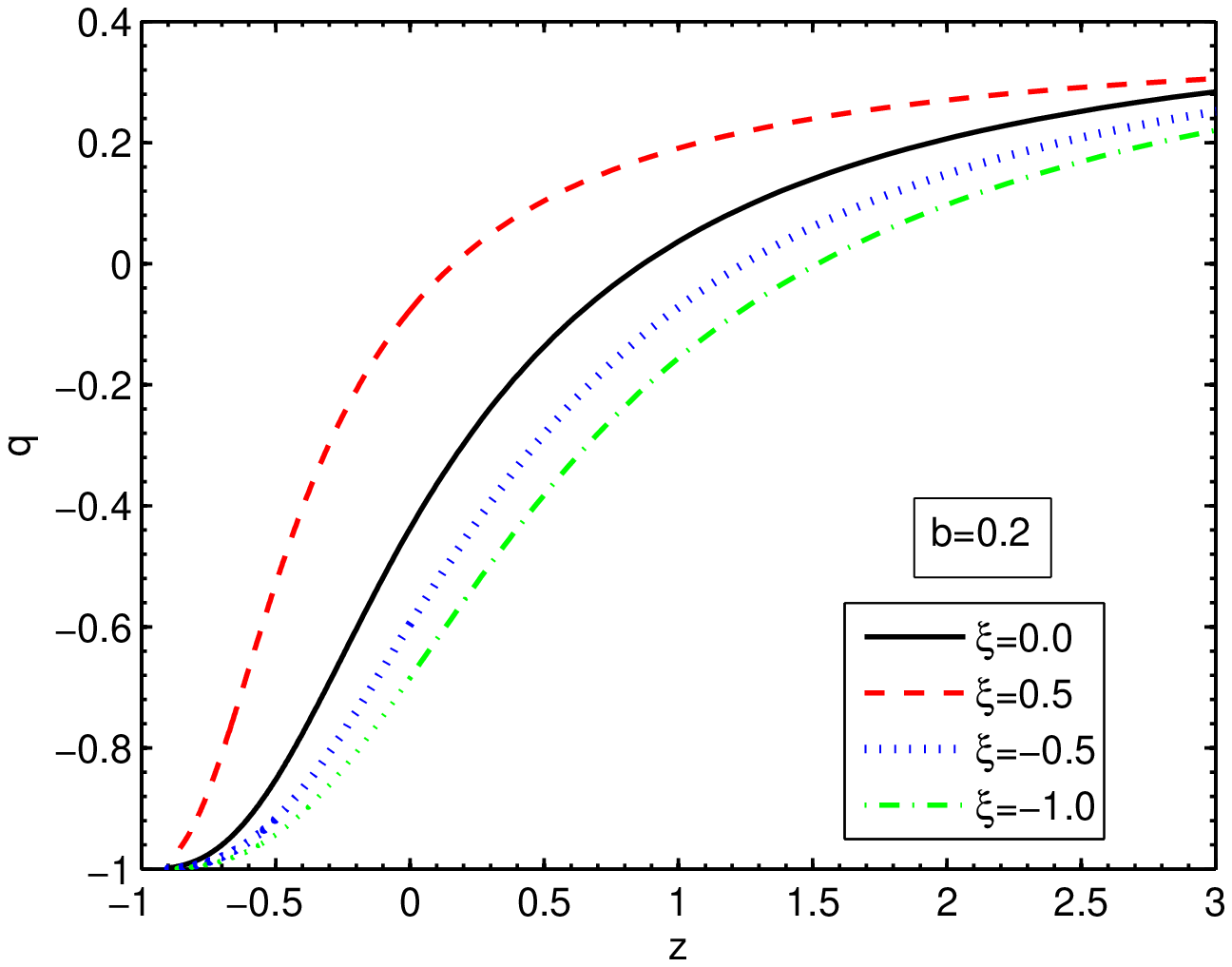}\includegraphics[width=8cm]{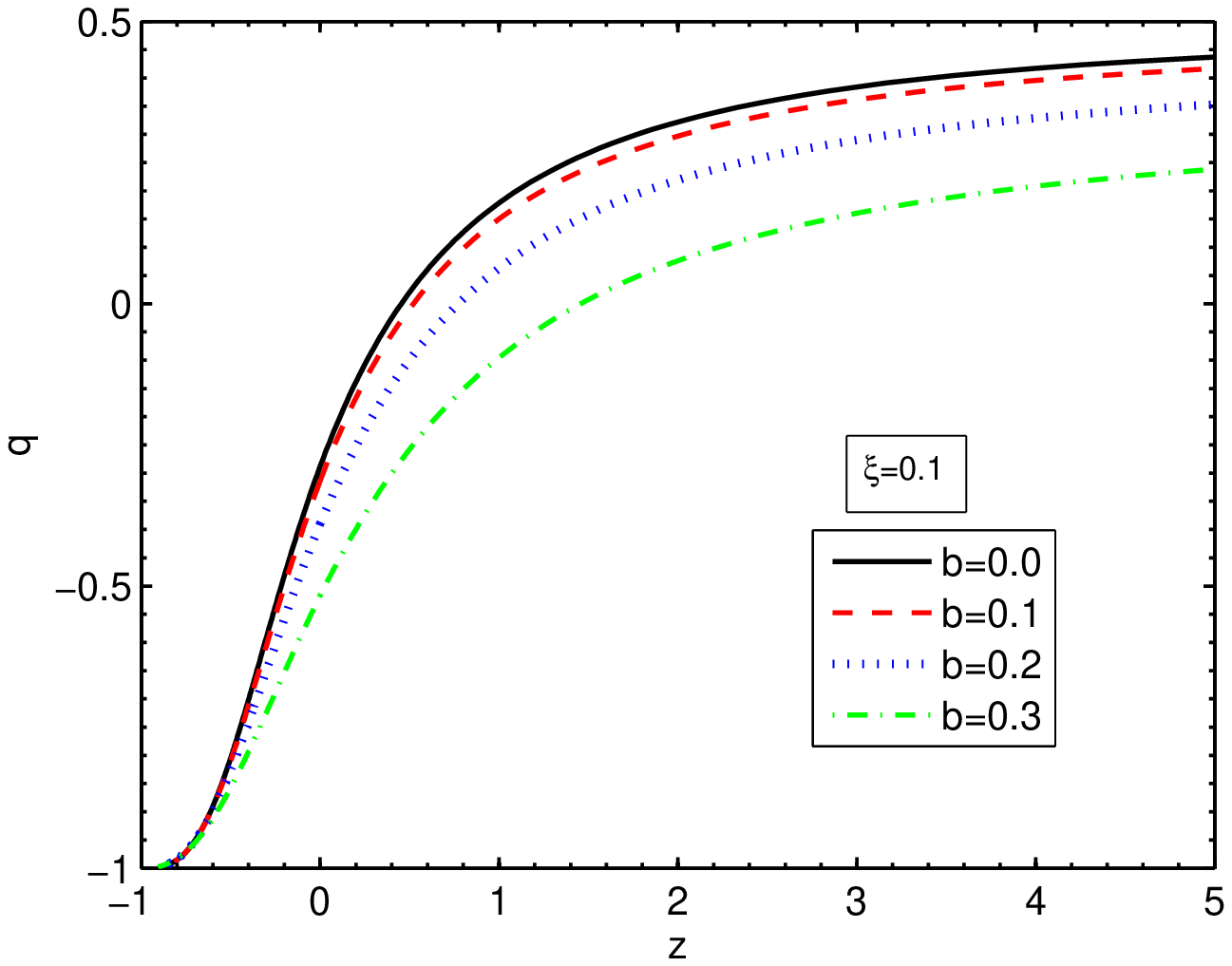}
\caption{The evolution of parameter $q$ for interacting case of GGDE
model as afunction of redshift $z$ as indicated in legend. }
\end{figure}

\begin{figure}[!htb]
\includegraphics[width=8cm]{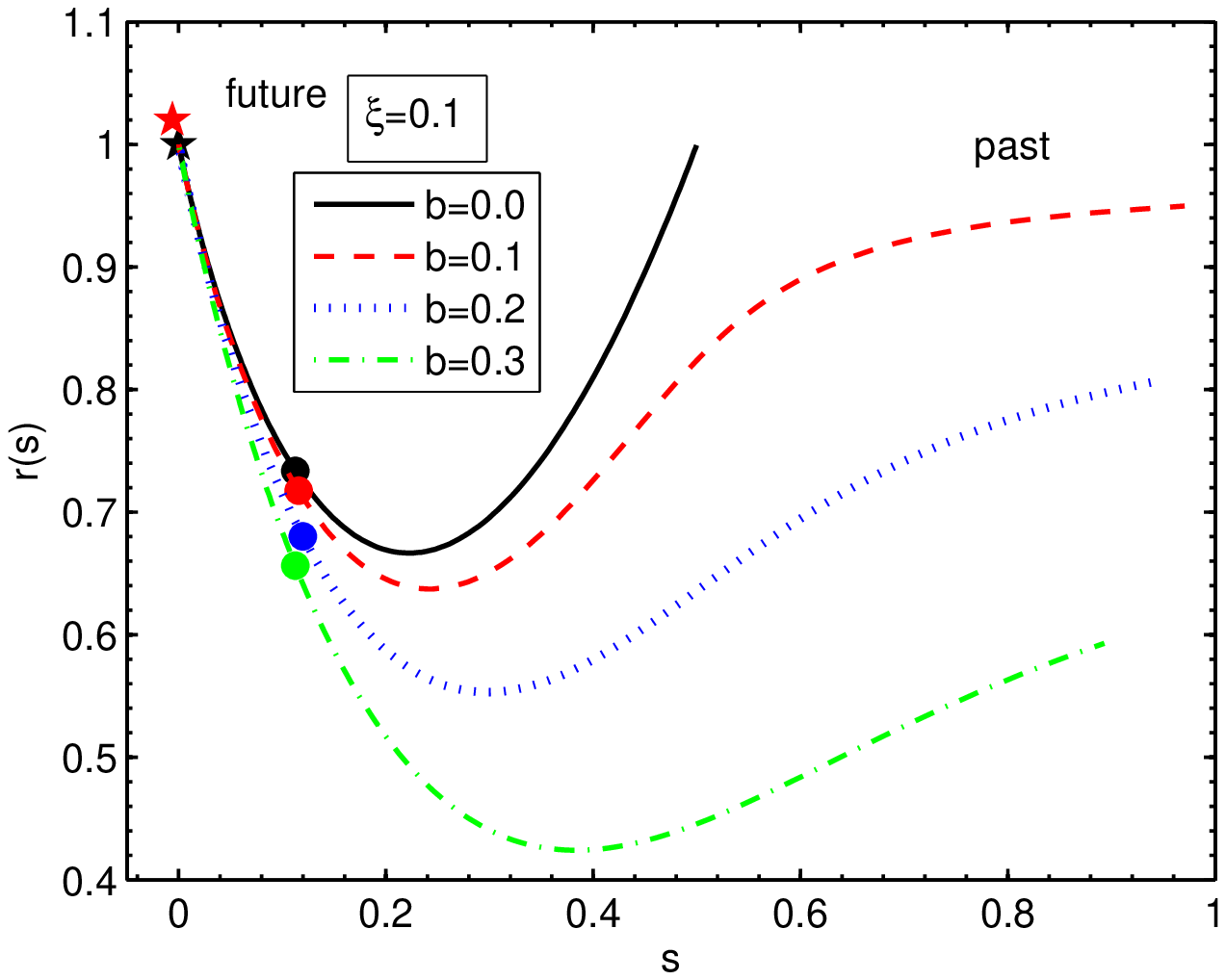}\includegraphics[width=8cm]{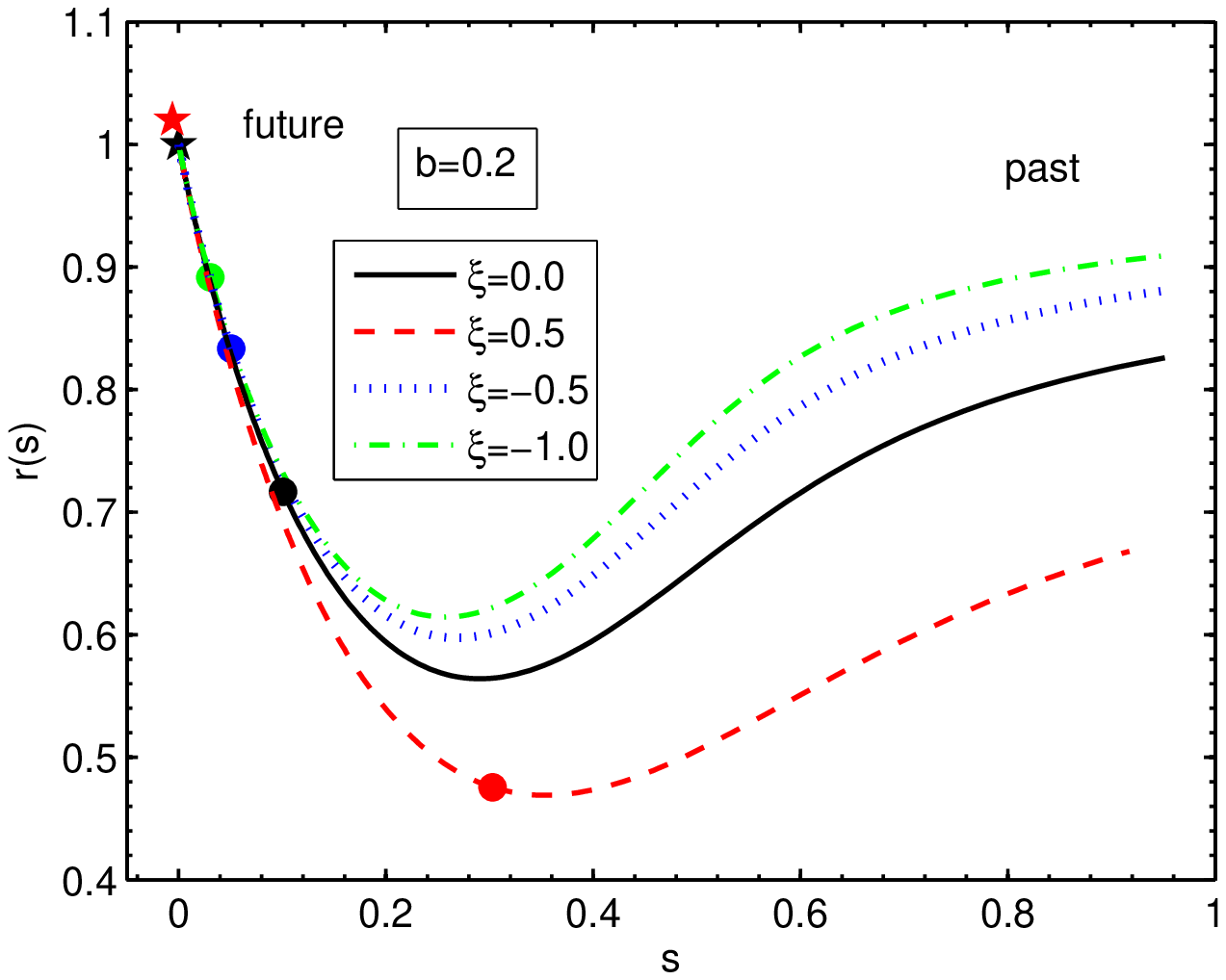}
\caption{The statefinder plot of interacting GGDE model. In left
panel the parameter $\xi$ is fixed and in right panel the parameter
$b$ is fixed. The present value $\{s_0,r_0\}$ is indicated by
colored circle on the curves. The location of $\Lambda$CDM model and
observational point are indicated by black and red stars,
respectively. }
\end{figure}

\begin{figure}[!htb]
\includegraphics[width=8cm]{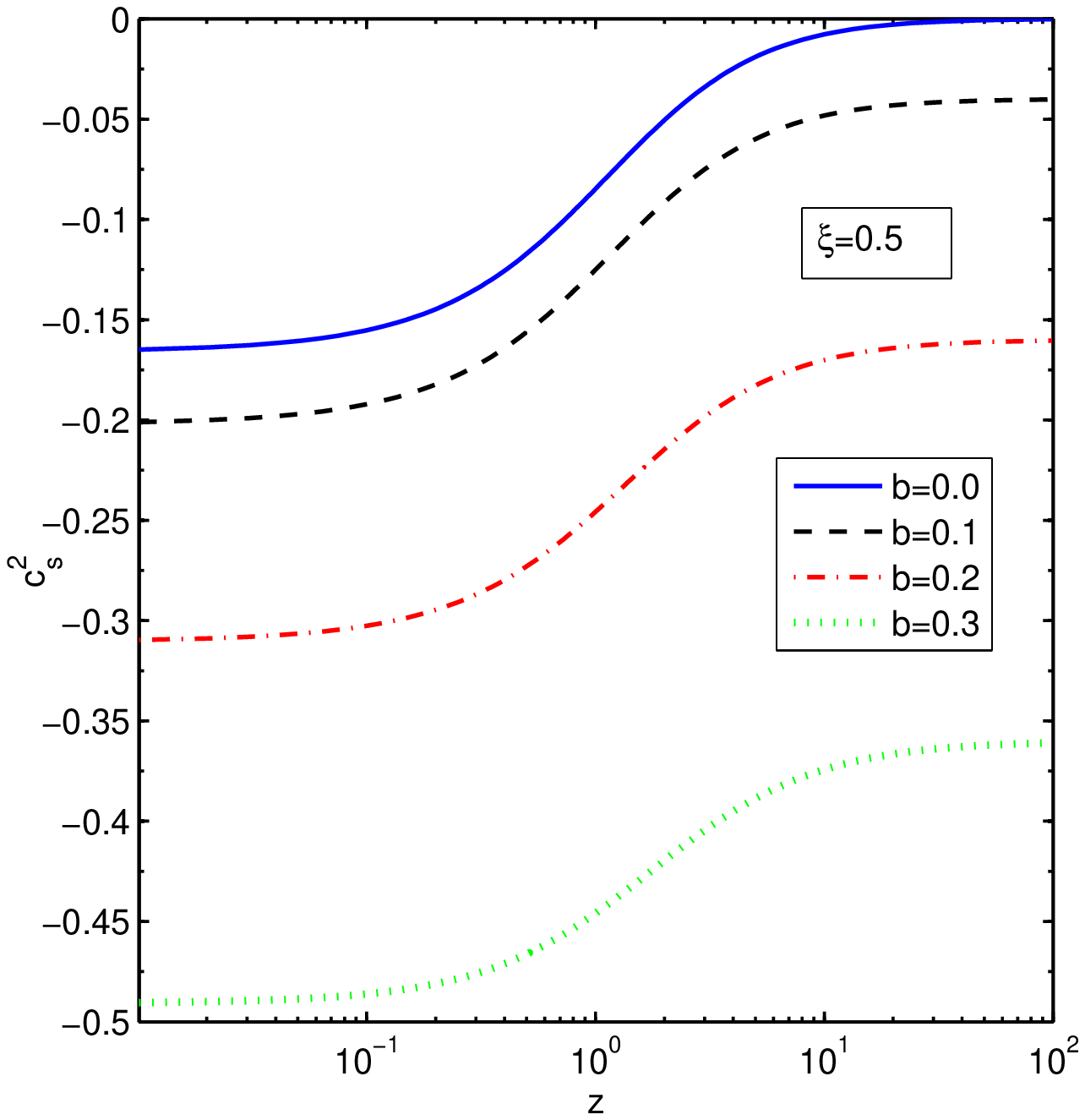}\includegraphics[width=8cm]{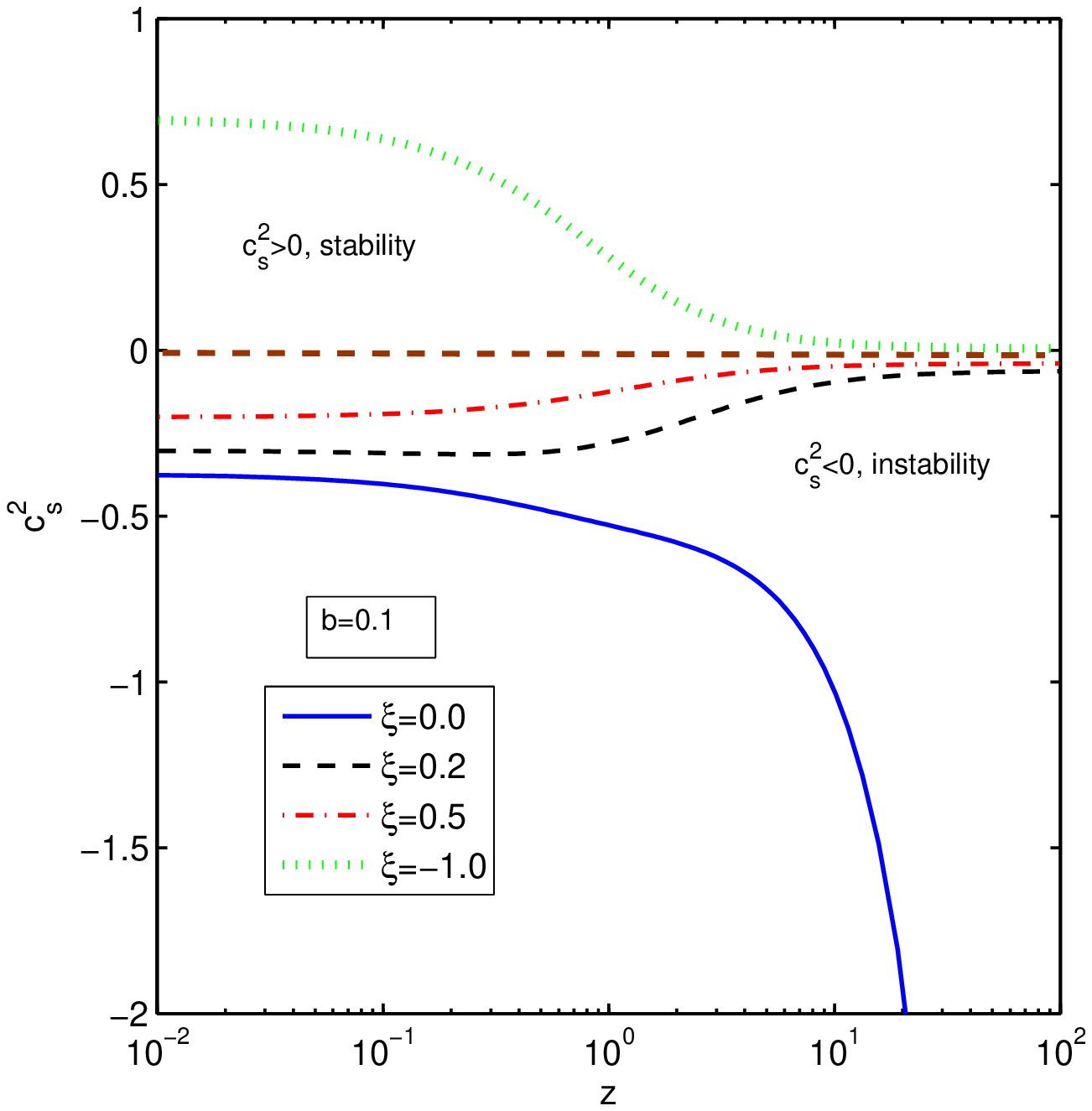}
\caption{ The adiabatic sound speed $c_s^2$ as a function of cosmic
redshift $z$ for different interaction parameter $b$ (left panel)
and various model parameter $\xi$ (right panel) as indicated in
legend. The horizontal dashed line in right panel separates the
stability and instability regions. }
\end{figure}


\newpage


\newpage
\begin{thebibliography}{99}
\bibitem{c1}
S. Perlmutter et al., Astrophys. J. {\bf 517}, 565 (1999).

\bibitem{c2}
C. L. Bennett et al., Astrophys. J. Suppl. {\bf 148}, 1 (2003).

\bibitem{c3}
M. Tegmark et al., Phys. Rev. D {\bf 69}, 103501 (2004).

\bibitem{c4}
S. W. Allen, et al., Mon. Not. Roy. Astron. Soc. {\bf 353}, 457
(2004).

\bibitem{winberg1}
S. Weinberg, Rev. Mod. Phys. 61 (1989) 1.

\bibitem{winberg2}
S. M. Carroll, Living Rev. Rel. 4 (2001) 1.

\bibitem{winberg3}
P. J. E. Peebles and B. Ratra, Rev. Mod. Phys. 75 (2003) 559.

\bibitem{winberg4}
T. Padmanabhan, Phys. Rept. 380 (2003) 235.

\bibitem{fin}
V. Sahni and A.A. Starobinsky, Int. J. Mod. Phys. D 9, 373 (2000).

\bibitem{reva}
E. J. Copeland, M. Sami and S. Tesujikawa, Int. J. Mod. Phys. D 15,
1753 (2006).

\bibitem{revb}
M. R. Setare, Phys. Lett. B \textbf{644}, 99, 2007.
\bibitem{revc}
M. R. Setare, Phys. Lett. B \textbf{654}, 1, 2007.
\bibitem{revd}
M. R. Setare, Phys. Lett. B \textbf{642}, 1, 2006.
\bibitem{reve}
M. R. Setare, Eur. Phys. J. C \textbf{50}, 991, 2007.
\bibitem{revf}
M. R. Setare, Phys. Lett. B \textbf{648}, 329, 2007.
\bibitem{revg}
M. R. Setare, Phys. Lett. B \textbf{653}, 116, 2007.

\bibitem{rev2}
M. Li, X. D. Li, S. Wang, Y. Wang, Commun. Theor. Phys. 56, 525,
2011.

\bibitem{rev3}
K. Bamba, S. Capozziello, S. Nojiri and S. D. Odintsov, Astrophys
Space Sci,(2012) 342, 155-228.

\bibitem{ghostq1}
F. R. Urban and A. R. Zhitnitsky, Phys. Lett. B 688 (2010) 9.

\bibitem{ghostq2}
 F. R. Urban and A. R. Zhitnitsky, Phys. Rev. D 80 (2009) 063001.

\bibitem{ghostq3}
F. R. Urban and A. R. Zhitnitsky, JCAP 0909 (2009) 018.

\bibitem{ghostq4}
F. R. Urban and A. R. Zhitnitsky, Nucl. Phys. B 835 (2010) 135.

\bibitem{ghostq5}
 A. R. Zhitnitsky, Phys. Rev. D 84, 124008 (2011).

\bibitem{ghostq6}
 N. Ohta, Phys. Lett. B 695, 41 (2011).

 \bibitem{ghostq7}
 R. G. Cai, Z. L. Tuo, H. B. Zhang and Q. Su, Phys. Rev. D 84, 123501
 (2011).

\bibitem{witq1}
 E. Witten, Nucl. Phys. B 156, 269 (1979).

\bibitem{witq2}
 G. Veneziano, Nucl. Phys. B 159, 213 (1979).

\bibitem{witq3}
 C. Rosenzweig, j. Schechter and C. G. Trahern, Phys. Rev. D 21, 3388
 (1980).

\bibitem{witq4}
  P. Nath and R. L. Arnowitt, Phys. Rev. D 23, 473 (1981).

\bibitem{witq5}
K. Kawarabayashi and N. Ohta, Nucl. Phys. B 175, 477 (1980).

\bibitem{witq6}
N. Ohta, Prog. Theor. Phys. 66, 1408 (1981).

\bibitem{zhiten1}
A. R. Zhitnitsky, Phys. Rev. D82, 103520 (2010).

\bibitem{zhinten2}
B. Holdom, Phys. Lett. B697, 351-356 (2011).

\bibitem{zhinten3}
E. Thomas, A. R. Zhitnitsky, Phys.Rev. D85 (2012) 044039.


\bibitem{caiq1}
R. G. Cai, Z. L. Tuo, H. B. Zhang and Q. Su, Phys. Rev. D 84, 123501
(2011).

\bibitem{ghostq22}
R. G. Cai, Z. L. Tuo, H. B. Zhang and Q. Su, Phys. Rev. D 84, 123501
(2011).

\bibitem{sheykhi1}
A. Sheykhi, M. S. Movahed, Gen. Relativ. Gravit. 44 (2012) 449.

\bibitem{sheykhi2a}
E. Ebrahimi and A. Sheykhi, Int. J. Mod. Phys. D20 (2011) 2369.

\bibitem{sheykhi2b}
 A. Sheykhi, M. Sadegh Movahed, E. Ebrahimi, Astrophys Space Sci 339
(2012)93.

\bibitem{sheykhi2c}
 A. Sheykhi, A. Bagheri, Euro. Phys. Lett., 95 (2011) 39001.

\bibitem{sheykhi2d}
 E. Ebrahimi and A. Sheykhi, Phys. Lett. B 706 (2011) 19.

\bibitem{sheykhi2e}
A. R. Fernandez, Phys.Lett.B 709 (2012) 313-321;


\bibitem{khod1}
A. Khodam-Mohammadi, M. Malekjani, M. Monshizadeh, Mod. Phys. Lett.
A, 27 (2012) 1250100.

\bibitem{malek1}
M. Malekjani, A. Khodam-Mohammadi,  Astrophys Space Sci. (2013) 343,
451-461.

\bibitem{zhit}
A. R. Zhitnitsky,  Phys. Rev. D 84, 124008 (2011).

\bibitem{h2}
M. Maggiore, L. Hollenstein, M. Jaccard and E. Mitsou, Phys. Lett. B
704, 102 (2011).

\bibitem{caighost}
R. G. Cai, Z. L. Tuo, Y. B. Wu, Y. Y. Zhao,     Phys.Rev. D86 (2012)
023511.

\bibitem{sheykhi3}
E. Ebrahimi, A. Sheykhi, arXiv:1209.3147, (2012).

\bibitem{sheykhi4}
A. Sheykhi, E. Ebrahimi, Y. Yosefi, arXiv:1210.0781, (2012).

\bibitem{sheykhib}
 Esmaeil E. Ebrahimi, A. Sheykhi, arXiv:1211.2686, (2012).

\bibitem{DM}
A. Bosma, Astron. J. 86 (1981) 1825.

\bibitem{DM2}
  P. J. E. Peebles, The Large-Scale Structure of the Universe (Princeton University Press,
NJ, 1980).

\bibitem{DM3}
  Padmanabhan, T., 1993. Structure Formation in the Universe. Cambridge University
  Press.

\bibitem{DM4}
 M. Malekjani , S. Rahvar , D.M.Z. Jassur, New Astronomy 14 (2009)
 398–405.

 \bibitem{DM5}
 M. Malekjani, H. Haghi, D. Mohammad-Zadeh Jassur, New Astronomy 17 (2012)
 149–153.

 \bibitem{interact}
 O. Bertolami , F. Gil Pedro and M. Le Delliou, Phys. Lett. B 654 (2007) 165.

 \bibitem{interact2}
 G. Olivares, F. Atrio, D. Pavon, Phys. Rev. D 71 (2005) 063523.

 \bibitem{feng}
 Feng, C., Wang, B., Gong, Y., Su, R.K.: J. Cosmol. Astropart. Phys.
0709, 005 (2007).

\bibitem{new1}
L. Amendola, Phys. Rev. D 62, 043511 (2000).

\bibitem{new2}
 L. Amen- dola and C.
Quercellini, Phys. Rev. D 68 (2003) 023514.

\bibitem{new3}
 L. Amendola, S.
Tsujikawa and M. Sami, Phys. Lett. B 632 (2006) 155.

\bibitem{new4}
 D. Pavon,W.
Zimdahl, Phys. Lett. B 628 (2005) 206.

\bibitem{new5}
S. Campo, R. Herrera, D. Pavon, Phys. Rev. D 78 (2008) 021302.

\bibitem{new6}
C. G. Boehmer, G. Caldera-Cabral, R. Lazkoz, R. Maartens, Phys. Rev.
D 78 (2008) 023505.
\bibitem{new7}
G. Olivares, F. Atrio-Barandela and D. Pavon, Phys. Rev. D 74 (2006)
043521.
\bibitem{new8}
S. B. Chen, B. Wang, J. L. Jing, Phys.Rev. D 78 (2008) 123503.

\bibitem{sah03}
Sahni, V., Saini, T.D., Starobinsky, A.A., Alam, U.: JETP Lett. 77,
201 (2003).

\bibitem{alamb}
Alam, U., Sahni, V., Saini, T.D., Starobinsky, A.A.: Mon. Not. R.
As- tron. Soc. 344, 1057 (2003)

\bibitem{kapa}
Capozziello, S., Cardone, V.F., Farajollahi, H., Ravanpak, A. arXiv:
1108.2789 (2011)

\bibitem{stateref1}
Zimdahl, W., Pavon, D., Gen. Relativ. Gravit. 36, 1483 (2004).

\bibitem{stateref2}
Zhang, X.: Phys. Lett. B 611, 1 (2005).

\bibitem{stateref3}
 Zhang, X.: Int. J. Mod. Phys. D 14, 1597 (2005).

\bibitem{stateref4}
 Setare, M.R., Zhang, J., Zhang, X.: J. Cosmol. Astropart. Phys.
0703, 007 (2007).

\bibitem{stateref5}
 Chang, B.R., Liu, H.Y., Xu, L.X., Zhang, C.W.,
Ping, Y.L.: J. Cosmol. Astropart. Phys. 0701, 016 (2007).

\bibitem{stateref6}
Malekjani, M., Khodam-Mohammadi, A. and N. Nazari-Pooya, Astrophys
Space Sci, 334:193–201, 2011.

\bibitem{stateref7}
 Zhang, L., Cui, J., Zhang, J., Zhang, X.: Int. J. Mod. Phys. D 19,21
 (2010).

\bibitem{stateref8}
Khodam-Mohammadi, A., Malekjani, M.: Astrophys. Space Sci. 331, 265
(2010).

\bibitem{stateref9}
 Wei, H., Cai, R.G.: Phys. Lett. B 655, 1 (2007).

\bibitem{stateref10}
Malekjani, M., Khodam-Mohammadi, A.: Int. J. Mod. Phys. D 19,1
(2010).
\bibitem{stateref11}
 Malekjani, M., Khodam-Mohammadi, A., Nazari-Pooya, N.,
Astrophys Space Sci (2011) 332, 515.

\bibitem{stateref12}
 Malekjani, M., Khodam-Mohammadi, A., Int. J. Theor. Phys. 51, 3141-3151
 (2012).
\bibitem{stateref13}
M. Malekjani · R. Zarei · M. Honari-Jafarpour, Astrophys Space Sci.
(2013) 346, 545-552.

\bibitem{malekghost}
M. Malekjani, A. Khodam-Mohammadi, Astrophys Space Sci (2013) 343,
451-461.

\bibitem{c21}
A. Sheykhi, Phys. Lett. B \textbf{680}, 113 (2009).

\bibitem{c21b}
 H. Wei \& R.
G. Cai, Phys. Lett. B \textbf{660}, 113 (2008).

\bibitem{c21c}
 L. Zhang, J. Cui, J. Zhang \& X. Zhang, Int. J. Mod. Phys. D \textbf{19}, 21 (2010).

\bibitem{obsegg1}
Alam, U., Sahni, V., Starobinsky, A.A.: J. Cosmol. Astropart. Phys.
06, 008 (2004).

\bibitem{obsegg2}
Huterer, D., Cooray, A.: Phys. Rev. D 71, 023506 (2005).

\bibitem{obsegg3}
Wang, Y., Tegmark, M.: Phys. Rev. D 71, 103513 (2005).

\bibitem{sound1}
 Y. S. Myung, Phys. Lett. B 652 (2007) 223.

\bibitem{sound2}
 K. Y. Kim, H. W. Lee and Y. S. Myung, Phys. Lett. B 660 (2008)
118. 115.

\end{thebibliography}
\end{document}